\documentclass[article,  twocolumn,showpacs,preprintnumbers,amsmath,amssymb]{revtex4}

\usepackage{graphicx}% Include figure files
\usepackage{dcolumn}% Align table columns on decimal point
\usepackage{bm}% bold math

\begin{document}
\author {Long Huang}\author{Xiaohua Wu}
\address{College of Physical Science and Technology, Sichuan University, 610064, Chengdu, China.}
\title{ A protocol to estimate the average fidelity of the bipartite system }
\begin{abstract}
   In the field of  quantum process tomography, the average fidelity quantifies how well the quantum channel preserves quantum information. In present work, we shall develop a protocol to estimate the average fidelity for the  bipartite system.
   We show that the average fidelity should be known if the
           three measurable quantities, the average survive probability of the product state and the average survive probability of each subsystem, have been decided. Our protocol can be also applied to decide the selected element of the quantum process matrix.

\end{abstract}
 \maketitle

\section{introduction}

  The characterization of the evolution
of a quantum system is one of the main tasks to accomplish to
achieve quantum information processing. A general class of methods,
which have been developed in quantum information theory to
accomplish this task is known as quantum process tomography (QPT)-
for a review of quantum tomography, see Refs. [$1-3$].
The various protocols of getting the complete information about the quantum process  can be divided into two classes: The  standard quantum process tomography (SQPT) [1,4,5], with the central idea of preparing  a set of linearly independent inputs and measuring the outputs via the quantum state tomography (QST), works without  requiring any additionally quantum resources.  Another is  the so-called
ancilla-assisted quantum process tomography (AAQPT) with  the approach of  encoding  all information about the transformation into a single bipartite
system-ancilla quantum state [6-8].

As a known fact,  the complete characterization of  a unknown quantum channel  is a
non-scalable task: For the  $N$ $d$-level system, there are about
$d^{4N}$ elements to be decided. Naturally, one may ask: Can the number of experiment
be scalable if partial information about the quantum process is to be characterized ?
 The average fidelity is an important quantity in QPT.
 It quantifies how well the quantum map preserves quantum information [9,10]. Recently, it has been shown that it is possible to estimate the average fidelity via a technique known as twirling [9-15]. The  Haar-twirl channel can be produced by the so-called Haar twirling procedure. It  consists  of applying  a  unitary, which is randomly chosen with the Haar measure,  before the process to be characterized, followed by the inverse of the  same unitary. The  average fidelity can be estimated by preparing an arbitrary pure state for the input of the Haar-twirl channel and then measuring its  survive probability ( the overlap between the input and output). Another twirling procedure, where the unitary is sampled uniformly from the Clifford group, has been proposed in [12]. The resulted Clifford-twirl channel was shown to be equivalent with
 the Haar-twirl one. Still, the  exact measurement of the average fidelity with the Clifford twirling protocol, which involves finite but exponentially large resources, is a non-scalable task. However, to experimentally characterize the fidelity of a quantum process on $n$ qubits
 for a desired accuracy, an efficient protocol has been constructed with quantum circuits of size $O(n)$ without requiring any ancilla qubits [12 ].

 Besides the twirling protocol, another important ancilla-less way for deciding the average fidelity has also been developed.  The average fidelity should be known if the survive probability of each state, which belongs to the state 2-design, has been decided [16-18].
 Furthermore, it was found that the state 2-design protocol can be also adapted to estimate an arbitrary element of the quantum process matrix.  For the $D$-dimensional system, the  state 2-design usually has more than $D^2$ elements in it. Therefore, an efficient method to estimate a selected element,  where the error scales as $\sqrt{1/M}$ with $M$ the number for the repetitions of the experiment, has also been proposed in [16-18].

The concept of the average fidelity can be generalized to the gate fidelity, a quantity characterizing how well the quantum map approximate a quantum gate [9].  It is demonstrated that twirling experiments previously used to characterize the average fidelity of quantum memories
efficiently can be easily adapted to estimate the average fidelity of the experimental implementation of important quantum computation processes,
such as untaries in the Clifford group, in a practical and efficient manner[19].

In present work, we shall develop a protocol to estimate the average fidelity of the quantum channel for a bipartite system. Our work is motivated by such an interesting case: In the Bell-type experiment, two spin-$s$ particles are initially prepared in an arbitrary state from a quantum source, then each particle is sent to Alice and  Bob, the two users who are space separated, respectively. In general, we suppose that there exists a quantum
map which relates the initial state (for the two particles in the source) to the final state (for the two particles in the users' hand).
Now, the average fidelity is still an important quantity to characterize  the  quantum process where the state should be kept unchanged.
For such a case, the average fidelity is hard to be measured in a directly way: By its definition, one should measure the survive probability of an arbitrary state. However, the problem appears when the two particles are prepared in an entangled state. As a solution for it, we introduce three directly measurable quantities, the average survive probability of the product state and the average survive probability of each subsystem, for the bipartite  system and give a formula to estimate  the average fidelity with the introduced quantities. Furthermore, we show that our protocol can be also applied to decide
an arbitrary selected element of the quantum process matrix.

The rest content of present work can be divided into following parts. In
Sec. II     we shall give a brief review of the known ancilla-less methods used to estimate the average fidelity of the quantum channel. Especially,
we introduce the convenient tool where a bounded matrix is related to a vector in the enlarged  Hilbert space. As an application of it, we show that
the average fidelity can be calculated as the expectation of the quantum process super operator with the separable Werner state. In Section III we shall firstly define the three quantities, the average survive probability of the product state and the average survive probability of each subsystem, for the bipartite system and then design several protocols  to measure them. A formula, where the average fidelity is related to the three average survive probabilities, should be constructed there. In Sec. IV we define the quantum process   matrix and its elements in an explicit way. By following the idea presented in [18], the protocol used to measure the average fidelity is adapted to decide an arbitrary element of the quantum process matrix. In Sec. V we shall develop an efficient protocol to measure the average fidelity. Finally, we end our work with a short discussion.

\section{ Measuring the average fidelity with the protocol of twirling }
In this section, we shall firstly give  a brief review  of the  known ancilla-less methods applied to measure the average fidelity of the quantum channel.  Let $\{\vert i\rangle\}_{i=1}^D$  the basis of a $D-$diamensional Hilbert space ${\mathrm{H}}_{D}$. For the state vector $\vert\psi\rangle=\sum_{i=1}^Dc_i\vert i\rangle$, the conjugated state vector $\vert \psi ^*\rangle $ is defined as $\vert\psi^*\rangle=\sum_{i=1}^Dc_i^*\vert i\rangle$. A corresponding capital letter, $\Psi$, is used to denote the projective operator, $\Psi=\vert\psi\rangle\langle\psi\vert$. With these denotations in hands, the average fidelity of the quantum
map $\varepsilon$ can be defined as,
\begin{equation}
f^{\mathrm{avg}}(\varepsilon)=\int d\mu_{\mathrm{H}}(\Psi)\mathrm{Tr}[\Psi\varepsilon(\Psi)],
\end{equation}
with $d\mu_{\mathrm{H}}(\Psi)$ the Haar-measure of states in $\mathrm{H}_D$ and   the process super operator $\varepsilon$    to be
$\varepsilon(\rho)=\sum_{n}A_n\rho A_n^{\dagger}$. Usually, we suppose $\varepsilon$ is trace preserving, $\sum _{n} A_{n}^{\dagger}A_{n}=\mathrm{I}_{D}$. As it is depicted in FIG. 1a, we  prepare an
arbitrary state $\vert \psi\rangle$ for input of the quantum channel $ \varepsilon$, after the evolution, measure the survival probability
$\mathrm{Tr}[\Psi\varepsilon(\Psi)]$. By sampling the state $\vert \psi\rangle$ with the Haar measure, the average fidelity of the channel $ \varepsilon$ should be decided.

Let $\vert \psi_0\rangle$ to be a fixed state in ${\mathrm{H}}_{D}$, one may relate the arbitrary state $\vert \psi\rangle $ to a unitary transformation $U$ ($U\in \mathrm{U}(D)$), $\Psi=U\Psi_{0}U^{\dagger}$. Now, the average fidelity in (1) can be rewritten as
\begin{equation}
f_b^{\mathrm{avg}}(\varepsilon)=\int d\mu_{\mathrm{H}}(U)\mathrm{Tr}[\Psi_{0}U^{\dagger}\varepsilon(U\Psi_0U^{\dagger})U].
\end{equation}
In the derivation of it, we have applied  the property  of the trace operation, $\mathrm{Tr}[ABC]=\mathrm{Tr}[BCA]$.
Here, we use a subscript $b$ to indicate that the average fidelity can be estimated with the Haar twirling protocol depicted in FIG. 1b: Apply a random unitary $U$ to the initial state $\vert \psi_0\rangle$, followed by the quantum operation $\varepsilon$, and then apply $U^{\dagger}$ to the output state. Then from (2), the average fidelity can be estimated by repeating the procedure with $U$ sampled randomly from the Haar measure in each experiment.

\begin{figure} \centering
\includegraphics[scale=0.4]{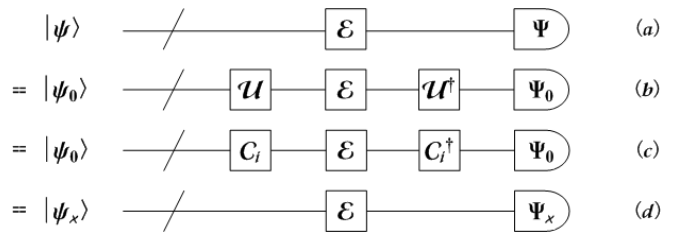}
\caption{\label{fig:epsart} (a)The average fidelity involves measuring the survive probability of an arbitrary state sampled with the Haar measure. It can be estimated with different protocols: (b) The Haar twirling, (c) the Clifford twirling, and (d) the sate 2-design.}
\end{figure}

In general, one may view the Haar twirling procedure as to prepare a so-called Haar-twirl channel $\varepsilon^{\mathrm{HT}}$,
\begin{eqnarray}
\varepsilon^{\mathrm{HT}}(\rho)&=&\int d\mu_{\mathrm{H}}(U) \mathcal{U}^{\dagger}\circ\varepsilon\circ \mathcal{U}(\rho)
\nonumber\\ &=&\int d\mu_{\mathrm{H}}(U)
U^{\dagger}\varepsilon(U\rho U^{\dagger})U].\nonumber
  \end{eqnarray}
In this picture, the average fidelity in (2) can be interpreted as  the survive probability of the fixed state $\Psi_0$ in the Haar-twirl channel
 $\varepsilon^{\mathrm{HT}}$, $f_b^{\mathrm{avg}}(\varepsilon)=\mathrm{Tr}[\Psi_0\varepsilon^{\mathrm{HT}}(\Psi_0)]$.

The Haar twirling is hard to realize in experiment since that it involves preparing a continuous  set of unitary operations.
To alleviate it,   the so-called Clifford-twirl channel $\varepsilon^{\mathrm{CT}}$ was introduced in [12],
\begin{eqnarray}
\varepsilon^{\mathrm{CT}}(\rho)&=&\frac{1}{K}\sum_{i=1}^{K}(\mathcal {C}^{\dagger}_{i}\circ \varepsilon\circ \mathcal{C}_{i})(\rho) \nonumber\\
&=&\frac{1}{K}\sum_{i=1}^{K}C_i^{\dagger}\varepsilon( C_i \rho C_{i}^{\dagger}) C_i,\nonumber
\end{eqnarray}
where $C_i$ are the elements of the Clifford group of ${\mathrm{H}}_{D}$.
To carry out the Clifford twirling in experiments, only a finite number ($K$) of operations should be prepared. It has been proven that the two different twirling procedures should result the same channel $\varepsilon^{\mathrm{CT}}=\varepsilon^{\mathrm{HT}}$. Certainly, as it is shown in FIG. 1c, the Clifford twirling protocol can be also applied to get the average fidelity. Formally, we express it as
\begin{eqnarray}
f_{c}^{\mathrm{avg}} (\varepsilon)=\frac{1}{K}\sum_{i=1}^{K}\mathrm{Tr}[\Psi_0C_{i}^{\dagger} \varepsilon(C_i\Psi_0 C_{i}^{\dagger})
C_i ].
\end{eqnarray}

Recently,  it has been found that a state  2-design can be also applied to measure the average fidelity. The state 2-design, $\{\Psi_x\}_{x=1}^{N}$, is set of
states satisfying the constraint that
   \begin{equation}
   \frac{1}{N}\sum_{x=1}^N \Psi_x\otimes \Psi_x=\frac{1}{D(D+1)}\sum_{i,j=1}^{D}\vert jk\rangle \langle jk\vert +\vert jk \rangle \langle kj\vert.  \end{equation}
Straitly to say, this definition is suitable for the case where all the  states $\Psi_x$ are equal weighted. For a more general definition of the state 2-design, please see [20].

For the case $N=D(D+1)$, the state 2-design is known to be a complete set of mutually unbiased bases (MUBs)[21-22]: That is a set of $D+1$
bases for ${\mathrm{H}}_{D}$ with a constant overlap of $1/D$ between elements of different bases,
\begin{equation}
\vert \langle \psi_{m}^{j}\vert \psi^{j'}_{m'}\rangle\vert^2=\{\begin{array}{c}
                                                                           \delta_{mm'}~~j=j' \\
                                                                           \frac{1}{D}~~~~~j\neq j'.
                                                                         \end{array}
\end{equation}

If $N=D^2$, the state 2-design is unique: It's just the symmetric information complete (SIC) set $\{\vert \psi^{\mathrm{SIC}}_x\rangle\}_{x=1}^{D^2}$ introduced in [23]. The normalized states $\vert \psi^{\mathrm{SIC}}_x\rangle$ have the property that
\begin{equation}
\vert\langle \psi^{\mathrm{SIC}}_x\vert \psi^{\mathrm{SIC}}_y\rangle\vert^2=\frac{1+D\delta_{xy}}{1+D}.
\end{equation}

The way of applying the state 2-design for estimate the average fidelity is shown in FIG. 1d: Preparing a state $\vert \psi_x\rangle$, which belongs to a given set of a state 2-design,   for the quantum channel $\varepsilon$, after the evolution, one  measure the expectation value of the projective operator $\Psi_x$ with the output $\varepsilon(\Psi_{x})$. The average fidelity should be known by repeating this process with a number of $N$ different inputs,
\begin{equation}
f_{d}^{\mathrm{avg}}( \varepsilon)=\frac{1}{N}\sum_{x=1}^{N}\mathrm{Tr}[\Psi_{x}\varepsilon(\Psi_{x})].
\end{equation}

In the above argument, we have assumed that all the quantities $f_k^{\mathrm{avg}}(\varepsilon)$, which are measured with the protocols depicted  from FIG. 1a to FIG. 1d, should equal the average fidelity defined in (1),
\begin{equation}
f^{\mathrm{avg}}_k ( \varepsilon)\equiv f^{\mathrm{avg}}(\varepsilon), ~~k=a,b,c,d
\end{equation}
(Here, we suppose $f^{\mathrm{avg}}_a ( \varepsilon)\equiv f^{\mathrm{avg}}(\varepsilon)$.)  Although   this  equivalence has been verified  in previous works, for the convenience of reading, we would still like to give a self -contained
 proof for it. To complete this task, we shall at first introduce the convenient tool where a bounded matrix in $\mathrm{H}_{D}$ is related to a  vector in the enlarged Hilbert space $\mathrm{H}_{D}^{\otimes 2}$. Let $A$ to be a bounded matrix in the $D-$ dimensional Hilbert space $\mathrm{H}_D$, with $A_{ij}=\langle i\vert  A\vert j\rangle$
the matrix elements for it, an isomorphism between  $A$ and   a $D^2-$dimensional
 vector $\vert A\rangle\rangle$   is defined as
 \begin{equation}
 \vert A\rangle\rangle =\sqrt{D} A\otimes \mathrm{I}_{D}\vert S_+\rangle=\sum_{i,j=1}^D A_{ij}\vert ij\rangle,
  \end{equation}
in which  $\vert S_+\rangle$ is the maximally entangled state for $\mathrm{H}_{D}^{\otimes 2}$, $\vert S_+ \rangle =\frac{1}{\sqrt{D}}\sum_{k=1}^{D}\vert kk\rangle$ with $\vert ij\rangle=\vert i\rangle\otimes \vert j\rangle$.
This   isomorphism   offers a one-to-one mapping between the matrix and its vector form.
  Suppose that  $A$ , $B$, and $\rho$  are three arbitrary bounded matrices in $\mathrm{H}_{D}$, there should be
\begin{equation}
\mathrm{Tr}[A^{\dagger}B]=\langle\langle A\vert B\rangle\rangle,
\vert A\rho B\rangle\rangle =A\otimes B^{\mathrm{T}}\vert \rho \rangle\rangle,
\end{equation}
with $B^{\mathrm{T}}$ denoting the transpose of B. Especially, if $A$ takes the form $A=\vert\psi\rangle\langle \phi\vert$, its corresponding vector should be
\begin{equation}
\vert  \vert\psi\rangle\langle \phi\vert\rangle\rangle= \vert \psi\rangle\otimes\vert \phi^*\rangle.
\end{equation}

With the isomorphism in (9) and its properties in (10-11), recalling  $\varepsilon(\rho)=\sum_{n}A_n\rho A_n^{\dagger}$, we are able to express the average quantities measured in FIG. 1 in the way like
\begin{equation}
f^{\mathrm{avg}}_k(\varepsilon )=\mathrm{Tr}[\hat{F}_k (\sum _{n}A_n\otimes A_n^*)],~~ k=a, b, c, d,
\end{equation}
 where  $\hat{F}_{k}$, the  super operators in $\mathrm{H}_{D}^{\otimes 2}$, are defined as
 \begin{eqnarray}
 \hat{F}_a &=&\int d\mu_{\mathrm{H}}(\Psi)\vert \Psi\rangle\rangle\langle\langle \Psi\vert, \\
 \hat{F}_b &=&\int d\mu_{\mathrm{H}}(U)U\otimes U^*\vert \Psi_0\rangle\rangle\langle\langle \Psi_0\vert (U\otimes U^*)^{\dagger}, \\
 \hat{F}_c &=& \frac{1}{K}\sum_{j=1}^{K}C_j\otimes C^*_j\vert \Psi_0\rangle\rangle\langle\langle \Psi_0\vert (C_j\otimes C_j^*)^{\dagger},\\
 \hat{F}_d &=&\frac{1}{N}\sum_{x=1}^{N}\vert \Psi_x\rangle\rangle\langle\langle \Psi_x\vert.
 \end{eqnarray}
 Now, if a separable Werner state $\rho^{\mathrm{sep}}_{\mathrm{W}}$ for $\mathrm{H}_{D}^{\otimes 2}$ is introduced as
 \begin{eqnarray}
 \rho^{\mathrm{sep}}_{\mathrm{W}}&=&\frac{1}{D(D+1)}(\mathrm{I}_D\otimes \mathrm{I}_D +D\vert S_+\rangle\langle S_+\vert)\nonumber\\
 &=&\frac{1}{D(D+1)}(\mathrm{I}_D\otimes \mathrm{I}_D +\vert\mathrm{I}_D\rangle\rangle\langle \langle\mathrm{I}_D\vert),
 \end{eqnarray}
 one may conclude that all the super operators $\hat{F}_k$ are equivalent since that
 \begin{equation}
 \hat{F}_k=\rho^{\mathrm{sep}}_{\mathrm{W}}.
 \end{equation}
 A simple reasoning,  where the above  conclusion can be  achieved at, is like this: At first, we take it for granted that $\hat{F}_a=\hat{F}_b$ since that the arbitrary state $\vert \psi\rangle$ in  FIG. 1a is related to the arbitrary unitary transformation $U$ in FIG. 1b via the simple relation,
 $\vert \psi\rangle= U\vert \psi_0\rangle$. At the same time, the relation $\hat{F}_b=\hat{F}_c$ should also hold because that both the Haar twirling and the Clifford twirling will result the same channel,  $\varepsilon^{\mathrm{CT}}=\varepsilon^{\mathrm{HT}}$. The proof for $\hat{F}_b=\rho^{\mathrm{sep}}_{\mathrm{W}}$ is given in Appendix. Recall that $\Psi_x$ is a Hermitian operator, $\Psi_x^*=\Psi^{\mathrm{T}}_x$. By performing the partial transposition on both sides of (4), the definition of the state 2-design may have an equivalent version:
 \[\frac{1}{N}\sum_{x=1}^{N}\Psi_x\otimes \Psi_x^*=\rho^{\mathrm{sep}}_{\mathrm{W}}.\]
Noting that $\Psi_x\otimes \Psi_x^*$ are product states, this is the reason why we call the Wernner state in (17) the separable one. From  (11), there   should be $ \vert\Psi_x\rangle\rangle\langle\langle\Psi\vert =\Psi_x\otimes \Psi_x^*$. Therefore,  the relation, $\hat{F}_d=\rho^{\mathrm{sep}}_{\mathrm{W}}$, can be easily verified.

\section{ estimating the average fidelity of the bipartite system}
In this section, we shall develop a protocol to estimate the fidelity of a bipartite system $\mathrm{H}=\mathrm{H}^A_D\otimes \mathrm{H}^B_D$. For this $D^2$-dimensional Hilbert system, we use $\Lambda$ to describe a trace preserving quantum map:
\[\Lambda(\rho)=\sum_{m}E_m\rho(E_m)^{\dagger}, \sum_{m}(E_m)^{\dagger}E_m=\mathrm{I}_D^{\otimes 2}.\]
For an arbitrary $\Lambda$, we can introduce the following three average quantities,
\begin{eqnarray}
\bar{f}_{AB}(\Lambda)=\int\int d\mu_{\mathrm{H}}(\Psi) d\mu_{\mathrm{H}}(\Phi) \mathrm{Tr} [\Psi\otimes \Phi\Lambda(\Psi\otimes \Phi)],  \\
\bar{f}_A(\Lambda)=\int\int d\mu_{\mathrm{H}}(\Psi) d\mu_{\mathrm{H}}(\Phi) \mathrm{Tr} [\Psi\otimes \mathrm{I}_D\Lambda(\Psi\otimes \Phi)],\\
\bar{f}_B(\Lambda)=\int\int d\mu_{\mathrm{H}}(\Psi) d\mu_{\mathrm{H}}(\Phi) \mathrm{Tr} [\mathrm{I}_D\otimes \Phi\Lambda(\Psi\otimes \Phi)],
\end{eqnarray}
where $\Psi\otimes \Phi$  denotes an arbitrary product state in $\mathrm{H}^A_D\otimes \mathrm{H}^B_D$ while  $d\mu_{\mathrm{H}}(\Psi)$ and  $d\mu_{\mathrm{H}}(\Phi)$ are the Haar measures of the states  on $\mathrm{H}_{D}$. In present work, $\bar{f}_{AB}(\Lambda)$ is referred to as the average survival
probability of the product states (for channel $\Lambda$),  $\bar{f}_A(\Lambda)$ and $\bar{f}_B(\Lambda)$ is the average survival probability
for subsystem system $\mathrm{H}_D^A$ and $\mathrm{H}^{B}_D$, respectively. An experimental protocol of measuring the above three quantities is
depicted in FIG. 2a: Preparing an arbitrary product state $\vert\psi\rangle\otimes \vert\phi\rangle$ as  the input for the quantum channel $\Lambda$, after the evolution, one may  simultaneously measure the three expectations with the output $\Lambda(\Psi\otimes \Phi)]$, $f_{AB}=\mathrm{Tr} [\Psi\otimes \Phi\Lambda(\Psi\otimes \Phi)]$,  $f_{A}=\mathrm{Tr} [\Psi\otimes \mathrm{I}_D\Lambda(\Psi\otimes \Phi)]$, and $f_B= \mathrm{Tr} [\mathrm{I}_D\otimes \Phi\Lambda(\Psi\otimes \Phi)]$. By sampling $\Psi$ and $\Phi$ randomly with the Haar measure, the average
survival probability in (19-21) should be known.

The average survive probabilities are defined in an integral version. To carry out the integrations, we shall also apply the isomorphism in (9). To let it has a form  suitable for the bounded operators in the  $D^2$-dimensional Hilbert space case, we introduce the following definition:
 Letting $\vert \Omega\rangle$ be a maximally entangled states in $\mathrm{H}^{\otimes 4}$,
$\vert \Omega\rangle=\frac{1}{D} \sum_{i,j=1}^D \vert ij ij\rangle$ with $\vert ijkl\rangle=\vert i\rangle\otimes \vert j\rangle\otimes\vert k\rangle\otimes \vert l\rangle$,
a vector $\vert  \Gamma)$ in $\mathrm{H}_{D}^{\otimes 4}$ is related to the  bounded operator $\Gamma$ in $H_D^{\otimes 2}$, with its matrix elements to be  $\Gamma_{ij;kl}\equiv \langle ij\vert \Gamma\vert kl\rangle$,  via the isomorphism,
\begin{equation}
\vert \Gamma)=D\cdot\Gamma\otimes \mathrm{I}_{D}^{\otimes 2}\vert\Omega\rangle=\sum_{i,j,k,l=1}^D \Gamma_{ij;kl}\vert ijkl\rangle.
\end{equation}
 Suppose that $\Gamma$, $\Delta$, and $\Sigma$ are three arbitrary bounded matrices in $\mathrm{H}_{D}^{\otimes 2}$, there should be
\begin{equation}
\mathrm{Tr}[ \Gamma^{\dagger}\Delta]=(\Gamma\vert\Delta),
\vert \Gamma\Sigma\Delta)=\Gamma\otimes \Delta^{\mathrm{T}} \vert \Sigma ).
\end{equation}
If $\Gamma=\vert \Psi\rangle\rangle\langle\langle \Phi\vert $, there exists such a relation,
\begin{equation}
\vert\vert \Psi\rangle\rangle\langle\langle \Phi\vert )=\vert \Psi\rangle\rangle\otimes \vert \Phi^*\rangle\rangle.
\end{equation}

\begin{figure} \centering
\includegraphics[scale=0.3]{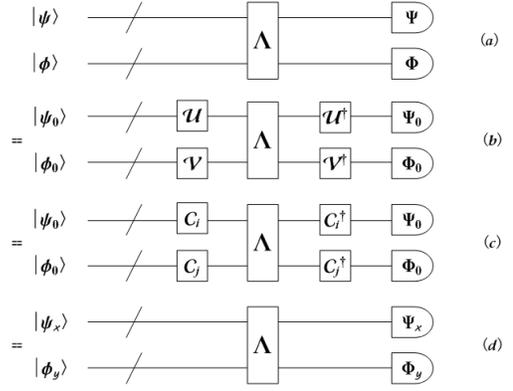}
\caption{\label{fig:epsart} (a) Circuit representation of measuring  the average survive probabilities  defined in (19-21). These  quantities can be also measured in different protocols: (b)The product
Haar twirling, (c) the product Clifford Twirling, and the product state 2-design.}
\end{figure}

As an application of the above isomorphism,  we find that the average survive probabilities in (19-21) can be rewritten as
\begin{eqnarray}
\bar{f}_{AB}(\Lambda)&=& \mathrm{Tr}[\hat{F}_{AB}\lambda],\\
\bar{f}_A(\Lambda)&=&\mathrm{Tr}[\hat{F}_A\lambda],\\
\bar{f}_B(\Lambda)&=&\mathrm{Tr}[\hat{F}_B\lambda],
\end{eqnarray}
where the four super operators in $H_{D}^{\otimes 4}$ are defined as
\begin{eqnarray}
\hat{F}_{AB}&=&\int\int d\mu_{\mathrm{H}}(\Psi) d\mu_{\mathrm{H}}(\Phi)\vert \Psi\otimes\Phi)(\Psi\otimes\Phi\vert,\\
\hat{F}_{A}&=&\int\int d\mu_{\mathrm{H}}(\Psi) d\mu_{\mathrm{H}}(\Phi) \vert \Psi\otimes\Phi)(\Psi\otimes \mathrm{I}_D\vert,\\
\hat{F}_{B}&=&\int\int d\mu_{\mathrm{H}}(\Psi) d\mu_{\mathrm{H}}(\Phi)\vert \Psi\otimes\Phi)(\mathrm{I}_D\otimes\Phi\vert, \\
\lambda&=&\sum_m E_m\otimes E_m^*.
\end{eqnarray}
Here, it should emphasize that $\lambda$ is a physical meaningful super operator: Suppose $\rho$ to be an arbitrary input for the quantum channel
$\Lambda$ and $\Lambda(\rho)=\sum_{m} E_m\rho E_m^{\dagger}$ to be the corresponding output.  Applying the result in (23),  we see that the two vectors, $\vert  \Lambda(\rho))$ and $\vert \rho)$, are simply related by $\lambda$,
\[\vert  \Lambda(\rho))=\vert \sum_{m} E_m\rho E_m^{\dagger})\equiv\lambda\vert \rho).\]

To carry out the integrations above, we introduce a special unitary transformation $\beta$ in $\mathrm{H}_D^{\otimes 4}$,
\begin{equation}
\beta=\sum_{i,j,k,l=1}^{D}\vert ijkl\rangle\langle ikjl\vert.
\end{equation}
One may check that $\beta$ is also a Hermitian operator,
\[\beta=\beta^{\dagger}=\beta^{-1}.\]
It has a nice property that
\[\beta\vert \Psi\otimes \Phi)=\vert \Psi\rangle\rangle\otimes \vert \Phi\rangle\rangle.\]
With the above property of $\beta$, we can reexpress $\hat{F}_{AB}$ as
\[ \hat{F}_{AB}=\beta(\int d\mu_{\mathrm{H}}(\Psi)\vert \Psi\rangle\rangle\langle\langle\Psi\vert \otimes \int d\mu_{\mathrm{H}}(\Phi)\vert \Phi\rangle\rangle\langle\langle \Phi\vert)\beta.\]
Recalling our results in (13-18), we have
\begin{equation}
\hat{F}_{AB}=\beta(\rho^{\mathrm{sep}}_{\mathrm{W}}\otimes\rho^{\mathrm{sep}}_{\mathrm{W}})\beta.
\end{equation}
Different protocols of measuring the $\bar{f}_{AB}(\Lambda)$ are depicted in FIG.2. We call the one in FIG. 2b as the product Haar twirling procedure: Let $\vert \psi_0\rangle$ and $\vert \Phi_0\rangle$ the fixed state for the subsystem $H_D^A$ and $H_D^{B}$, respectively. The two arbitrary states in (19), $\Psi$ and $\Phi$, may be related to the arbitrary unitary operation $U$ and $V$
\[\Psi=U\Psi_0U^{\dagger}, \Phi=V\Phi_0V^{\dagger},~~~U,V\in \mathrm{U}(D),\]
respectively.
The average survive probability of the product states may be expressed as
$\bar{f}_{AB}^b(\Lambda)=\int d\mu_{H}(U)\int d\mu_{H}(V)\mathrm{Tr}[\Psi_0\otimes \Phi_0 (\mathcal{U\otimes V })^{\dagger}\circ\Lambda\circ\mathcal{U\otimes V}(\Psi_0\otimes\Phi_0)]$. In experiment, we first prepare $\vert \psi_0\rangle\otimes\vert \phi_0\rangle $ as the fixed input, then apply
the operation $U\otimes V$ before the map $\Lambda$  and an operation $(U\otimes V)^{\dagger}$ after. Finally, we measure the survive probability of  $\vert \psi_0\rangle\otimes\vert \phi_0\rangle $ with the output. By sampling $U$ and $V$ randomly with the Haar measure of $\mathrm{U}(D)$, we shall get the quantity $\bar{f}_{AB}^b(\Lambda)$ defined above.

With the isomorphism in (22), one may express $\bar{f}_{AB}^b(\Lambda)$ as
\[\bar{f}_{AB}^b(\Lambda)=\mathrm{Tr}[\hat{F}^b_{AB}\lambda],\]
where  the super operator $\hat{F}^b_{AB}$ has the form
\[ \hat{F}^b_{AB}=\beta(\hat{F}_b\otimes\hat{F}_b)\beta \]
with $\hat{F}_b$ defined in (14). Applying the result in (18), we find $ \bar{f}_{AB}^b(\Lambda)$ equals the quantity $\bar{f}_{AB}(\Lambda)$. Therefore, the product Haar twirling procedure in FIG. 2b represents a possible way of getting $\bar{f}_{AB}(\Lambda)$.

The way of applying the product  Clifford twirling procedure to measure the average survive probability of the product states is  shown  in FIG. 2c.
Its experimental data can be collected as $\bar{f}_{AB}^c(\Lambda)=\frac{1}{K^2}\sum_{i,j=1}^{K}\mathrm{Tr}[\Psi_0\otimes \Phi_0 (\mathcal{C}_{i}\otimes \mathcal{C}_j)^{\dagger}\circ\Lambda\circ\mathcal{C}_i\otimes \mathcal{C}_j(\Psi_0\otimes\Phi_0)]$, where $\{C_i\}_{i=1}^K$ is the Clifford group of $\mathrm{H}_D$. Via the similar argument above, we have
 $\bar{f}_{AB}^c(\Lambda)=\mathrm{Tr}[\hat{F}^c_{AB}\lambda]$
with the super operator $\hat{F}^c_{AB}$ to be
$ \hat{F}^c_{AB}=\beta(\hat{F}_c\otimes\hat{F}_c)\beta .$
Obviously, $\bar{f}_{AB}^c(\Lambda)$ measured with the product Clifford twirling protocol  equals $\bar{f}_{AB}(\Lambda)$ in (19) since that
$\hat{F}^c_{AB}=\hat{F}_{AB}$.

Let $\{\Psi_x\}_{x=1}^{N}$ and $\{\Phi_y\}_{y=1}^N$  to be the state 2-designs defined in (4),
the quantity $\bar{f}_{AB}^d(\Lambda)$,
$\bar{f}_{AB}^d(\Lambda)=\frac{1}{N^2}\sum_{x,y=1}^{N}\mathrm{Tr}[(\Psi_x\otimes \Phi_y)\Lambda(\Psi_x\otimes \Phi_y)]$, can be directly measured through the  method in FIG. 2d.  Jointing the result    $\bar{f}_{AB}^d(\Lambda)=\mathrm{Tr}[\beta(\hat{F}_d\otimes \hat{F}_d)\beta \lambda]$ with (18) and (33), we conclude that $\bar{f}_{AB}^d(\Lambda)= \bar{f}_{AB}(\Lambda)$.

After giving a detail discussion about how to measure  $\bar{f}_{AB}(\Lambda)$ in experiment, we shall focus on $\bar{f}_{A}(\lambda)$ and $\bar{f}_{B}(\Lambda)$, the average survive probabilities for the subsystems, defined in (20) and (21), respectively. Let's consider $\bar{f}_{A}(\Lambda)$ at first. Using Shur's lemma,
\[\int d\mu_{\mathrm{H}}(\Phi)\Phi=\frac{1}{D}\mathrm{I}_D,\]
 we can simplify the expression of $\hat{F}_A$ in (29) as
 \[\hat{F}_A=\frac{1}{D}\int d\mu_{\mathrm{H}}(\Psi)  \vert \Psi\otimes\mathrm{I}_D)(\Psi\otimes \mathrm{I}_D\vert.\]
With  the unitary $\beta$ in (32), it has an equivalent form
\[\hat{F}_A=\beta(\int d\mu_{\mathrm{H}}(\Psi)\vert \vert\Psi\rangle\rangle\langle \Psi\vert \otimes \frac{\vert\mathrm{I}_D\rangle\rangle\langle\langle \mathrm{I}_D\vert}{D})\beta.\]
Recalling our definition of the super operator $\hat{F}_a$ in (13) and the result in (18),  we can get the formula
\begin{equation}
 \hat{F}_A=\beta(\rho_{\mathrm{W}}^{\mathrm{sep}}\otimes \frac{\vert \mathrm{I}_D\rangle\rangle\langle\langle \mathrm{I}_D\vert}{D})\beta.
 \end{equation}
Via a similar argument, there should be
\begin{equation}
 \hat{F}_B=\beta(\frac{\vert \mathrm{I}_D\rangle\rangle\langle\langle \mathrm{I}_D\vert}{D}\otimes\rho_{\mathrm{W}}^{\mathrm{sep}} )\beta.
 \end{equation}

With the  result in (18), where the different ways to expand the  separable Werner are given, we are able to prove that the average survive probability for a selected  subsystem can be measured  by the various protocols in FIG. 2. For example, when the survive  probability of subsystem $H^{\mathrm{A}}_{D}$ is measured with the product state 2-design protocol depicted in FIG. 2d, the experimental data can be organized in the way like
\begin{equation}
\bar{f}_A^d(\Lambda)=\frac{1}{N^2}\sum_{x,y=1}^N\mathrm{Tr}[(\Psi_x\otimes \mathrm{I}_D)\Lambda(\Psi_x\otimes\Phi_y)].
\end{equation}
Formally, it can be transferred  into $\bar{f}_A^d(\Lambda)=\mathrm{Tr}[\hat{F}_{A}^{d}\lambda]$ with the super operator
$\hat{F}_{A}^{d}$ to be
\[\hat{F}_{A}^{d}=\frac{1}{N^2}\sum_{x,y=1}^N\vert \Psi_x\otimes\Phi_y)(\Psi_x\otimes \mathrm{I}_D\vert.\]
As it is shown in [20], the state 2-design has the property that
\[\frac{1}{N}\sum_{y=1}^N\Phi_y=\frac{1}{D}\mathrm{I}_D.\]
With this property in hand, we rewrite $\hat{F}_{A}^{d}$ as
\[ \hat{F}_{A}^{d}=\beta[ \frac{1}{N}\sum_{x=1}^{N}(\Psi_x\rangle\rangle\langle\langle \Psi_x\vert)\otimes \frac{\vert \mathrm{I}_D\rangle\rangle\langle\langle   \mathrm{I}_D\vert }{D}
 ]\beta.\]
Jointing it with (16) and (18), we find $\hat{F}_{A}^{d}=\hat{F}_{A}$.   Therefore, $\bar{f}_A(\Lambda)$ can be measured in the way presented in (36).

With the separable Werner state defined in (17), we shall get a relation, $\mathrm{I}_D^{\otimes 2}= D(D+1)\rho_{\mathrm{W}}^{\mathrm{sep}}-\vert \mathrm{I}_D\rangle\rangle\langle\langle I_D\vert$. From it, we can expand the identity operator $\mathrm{I}_{D}^{\otimes 4}$ as
\begin{eqnarray}
\mathrm{I}_{D}^{\otimes 4}=D^2(D+1)^2\rho_{\mathrm{W}}^{\mathrm{sep}}\otimes \rho_{\mathrm{W}}^{\mathrm{sep}}
+\vert \mathrm{I}_D\rangle\rangle\langle\langle I_D\vert\otimes\vert \mathrm{I}_D\rangle\rangle\langle\langle I_D\vert \nonumber\\
-D(D+1)[\rho_{\mathrm{W}}^{\mathrm{sep}}\otimes \vert \mathrm{I}_D\rangle\rangle\langle\langle \mathrm{I}_D\vert+ \vert\mathrm{I}_D\rangle\rangle\langle\langle \mathrm{I}_D\vert\otimes \rho_{\mathrm{W}}^{\mathrm{sep}}]\nonumber
\end{eqnarray}
Note that $\mathrm{I}_{D}^{\otimes 4}$ is invariant under the transformation of $\beta$, $\mathrm{I}_{D}^{\otimes 4}=\beta\mathrm{I}_{D}^{\otimes 4}\beta$. Recalling our definitions of the super operators, $\hat{F}_{AB}$ in (33), $\hat{F}_{A}$ in (34), and $\hat{F}_{B}$ in (35), we  shall
find that the identity operator,  $\mathrm{I}_{D}^{\otimes 4}$,  can   also  be expanded as
\begin{equation}
 \mathrm{I}_{D}^{\otimes 4}=D^2(D+1)^2[\hat{F}_{AB}-\frac{\hat{F}_{A}+\hat{F}_{B}}{D+1}]+\vert \mathrm{I}_D^{\otimes 2})(\mathrm{I}_{D}^{\otimes 2}\vert.
 \end{equation}

For the quantum channel $\Lambda$, $\Lambda(\rho)=\sum_{m}E_m \rho E^{\dagger}_m$, of the joint system $\mathrm{H}=\mathrm{H}_{D}^A\otimes \mathrm{H}_{D}^B$,  we
introduce the entanglement  fidelity $ f^{\mathrm{ent}}(\Lambda)$, which has been proposed to  characterize  the noise strength in $\Lambda$ [24], in the way like
\begin{eqnarray}
f^{\mathrm{ent}}(\Lambda)
&=&\langle\Omega\vert( \mathrm{I}_D^{\otimes 2}\otimes \Lambda)(\vert\Omega\rangle\langle\Omega\vert)\vert \Omega\rangle \nonumber\\
&=&\frac{1}{D^4}\sum_{m}\vert \mathrm{Tr}[E_m]\vert^2,
\end{eqnarray}
where  $\vert \Omega\rangle$, as it has been introduced  in (22),  is the maximally entangled state of $\mathrm{H}\otimes \mathrm{H}^{\mathrm{anc}}$ with $\mathrm{H}^{\mathrm{anc}}$ to be a $D^2$-dimensional ancilla system. With the super operator $\lambda$ in (31), one may express the entanglement fidelity
with a more compact form:
\begin{equation}
 f^{\mathrm{ent}}(\Lambda)=\frac{1}{D^4}\mathrm{Tr}[\lambda].
 \end{equation}
Jointing it with our expanding of the identity operator in (37) and the equations (25-27), we find that the entanglement fidelity should be known
if the three average survive probabilities, $\bar{f}_{AB}(\Lambda)$, $\bar{f}_{A}(\Lambda)$ and $\bar{f}_{B}(\Lambda)$, have been decided,
\begin{equation}
f^{\mathrm{ent}}(\Lambda)=\frac{1}{D^2}\{1+(D+1)^2\bar{f}_{AB}-(D+1)[\bar{f}_{A}+\bar{f}_{B}]\}.
\end{equation}
(For simplicity, we  have omitted the symbol  $(\Lambda)$ in the expression for each average survive probability.)
In the derivation of it, we have used the formula
\begin{equation}
(\mathrm{I}_{D}^{\otimes 2}\vert \lambda =\sum_{m}(E_m^{\dagger}E_m\vert=(\mathrm{I}_{D}^{\otimes 2}\vert,
\end{equation}
which can be viewed as a vector-form expression of trace-preserving condition of $\Lambda$.

Now, we are able to show that the average fidelity can also be estimated with the three quantities, $\bar{f}_{AB}(\Lambda)$, $\bar{f}_{A}(\Lambda)$ and $\bar{f}_{B}(\Lambda)$. As a well-known fact [9,10], there exists a beautiful formula where the  average fidelity is  simply related to the entanglement fidelity. Instead of directly citing this formula,   we shall give  a simple reasoning to recover it.
We use $\tilde{\rho}^{\mathrm{sep}}_{\mathrm{W}}$ as the generalization of the separable Werner state (for the $D^2$-dimensional system) in (17)  for the $D^4$-dimensional system,
\begin{equation} \tilde{\rho}^{\mathrm{sep}}_{\mathrm{W}}=\frac{1}{D^2(D^2+1)}(\mathrm{I}_D^{\otimes 4}+\vert \mathrm{I}_D^{\otimes 2})(\mathrm{I}_D^{\otimes2} \vert).
\end{equation}
Correspondingly, the results in (12) and (18) can also generalized into the following form
\[f^{\mathrm{avg}}(\Lambda)=\mathrm{Tr}[\tilde{\rho}^{\mathrm{sep}}_{\mathrm{W}}\lambda].\]
By jointing it with (39) and (41), the well-known relation, for the case where the process super operator  $\Lambda$ is defined for the $D^2$-dimensional system, is recovered  here
\begin{equation}
f^{\mathrm{avg}}(\Lambda)=\frac{D^2f^{\mathrm{ent}}(\Lambda)+1}{D^2+1}.
\end{equation}
Putting (40) into it, we shall arrive at one of the main results of present work,
\begin{equation}
f^{\mathrm{avg}}(\Lambda)=\frac{1}{D^2+1}\{2+(D+1)^2\bar{f}_{AB}-(D+1)[\bar{f}_{A}+\bar{f}_{B}]\},
\end{equation}
where we show that the average fidelity $f^{\mathrm{avg}}(\Lambda)$ can be estimated with the average survival probabilities, $\bar{f}_{AB}(\Lambda)$, $\bar{f}_{A}(\Lambda)$ and $\bar{f}_{B}(\Lambda)$.

\section{ measuring the selected element of the process matrix}

The proposal,  which suggests that  the protocol designed for measuring the average fidelity can be also applied to decide an arbitrary element of the quantum process matrix, originated from the work in [16]. In the first version of it, the off-diagonal elements should be measured by introducing ancilla system. Recently, an improved scheme, where all the elements  can be estimated without introducing any ancilla system, has been presented in
[18]. In the argument below,  following  the idea presented in [18], we shall develop a method where an arbitrary element of the process matrix can be decided with the average survive probabilities.

At beginning, let's introduce the definition of the process matrix.  For a reason which will be clear later, we suppose   $\bar{\Lambda}$ to be an arbitrary  quantum map for the joint system $\mathrm{H}=\mathrm{H}_{D}^A\otimes \mathrm{H}_{D}^B$,
\begin{equation}
\bar{\Lambda}(\rho)=\sum_{n}\bar{E}_{n}\rho \bar{E}_{n}, \sum_{n}\bar{E}_{n}^{\dagger}\bar{E}_n=\mathrm{I}_{D}^{\otimes 2}.
\end{equation}
Letting $\{\Gamma_{\mu}\}_{\mu=1}^{D^4}$ to be  an orthogonal  operator basis for $\mathrm{H}$,
\begin{equation}
\mathrm{Tr}[\Gamma^{\dagger}_{\mu}\Gamma_{\nu}]=D^2\delta_{\mu\nu},
\end{equation}
one may rewrite (45) as
\begin{equation}
\bar{\Lambda}(\rho)=\sum_{\mu,\nu=1}^{D^4} \Gamma_{\mu}\rho \Gamma_{\nu}^{\dagger}\chi_{\mu;\nu}(\bar{\Lambda}),
\end{equation}
where the coefficients $\chi_{\mu;\nu}(\bar{\Lambda})$,
\begin{equation}
\chi_{\mu;\nu}(\bar{\Lambda})=\frac{1}{D^4}\sum_{n}\mathrm{Tr}[\Gamma_{\mu}^{\dagger} \bar{E}_n](\mathrm{Tr}[\Gamma_{\nu}^{\dagger} \bar{E}_n])^*,
\end{equation}
are the entries of a $D^4\times D^4$ process matrix $\chi(\bar{\Lambda})$ which is Hermitian by definition. Here, it should be noted the factor $D^{-4}$ comes from the fact that the  basis operators in (46) are not normalized.  As it has been done in previous works,
we suppose that $\Gamma_{\mu}$ are Hermitian and unitary operators,
\[\Gamma_{\mu}=\Gamma_{\mu}^{\dagger}=\Gamma_{\mu}^{-1}.\]
Furthermore, we rewrite  (48) with a convenient form
\begin{equation}
\chi_{\mu;\nu}(\bar{\Lambda})=\frac{1}{D^4}\mathrm{Tr}[(\sum_{n}\bar{E}_n\otimes \bar{E}_n^*)(\Gamma_{\mu}\otimes \Gamma_{\nu}^*)].
\end{equation}
From it, we see that the diagonal matrix elements should take the form,
 \begin{equation}
\chi_{\mu;\mu}(\bar{\Lambda})=\frac{1}{D^4}\mathrm{Tr}[(\sum_{n}\bar{E}_n\otimes \bar{E}_n^*)(\Gamma_{\mu}\otimes \Gamma_{\mu}^*)].
\end{equation}
Now, let's assume that the quantum map $\Lambda$ discussed in above section  is related to $\bar{\Lambda}$ via the simple form,
\[\Lambda=\bar{\Lambda} \circ {\Gamma}_{\mu}, \]
where  the process super operator $\bar{\Lambda} \circ {\Gamma}_{\mu}$ is defined as $\bar{\Lambda} \circ {\Gamma}_{\mu}(\rho)=\sum_n \bar{E}_n(\Gamma_{\mu}\rho\Gamma_{\mu}^{\dagger})\bar{E}_n^{\dagger}$. We call the so-defined $\Lambda$ the modified map (of $\bar{\Lambda}$).
Based on this assumption, we find that the supper operator $\lambda$ in (31)  should be
\[\lambda=(\sum_{n}\bar{E}_n\otimes \bar{E}_n^*)(\Gamma_{\mu}\otimes \Gamma_{\mu}^*).\]
Recalling the formula for the calculation of the entanglement fidelity in (39) and the one for the diagonal matrix elements in (50), we shall find an interesting result,
\begin{equation}
\chi_{\mu;\mu}(\bar{\Lambda})=f^{\mathrm{ent}}(\bar{\Lambda} \circ {\Gamma}_{\mu})\equiv f^{\mathrm{ent}}(\Lambda).
\end{equation}
In other words, one may transfer the task of measuring the diagonal matrix element, which is defined for the quantum map $\bar{\Lambda}$, into the
one of deciding the entanglement fidelity of the modified map $\Lambda=\bar{\Lambda} \circ {\Gamma}_{\mu}$.
Based on this principle,  the diagonal matrix element $\chi_{\mu;\mu}(\bar{\Lambda})$ can be estimated with $\bar{f}_{AB}(\bar{\Lambda}\circ {\Gamma}_{\mu})$,
$\bar{f}_{A}(\bar{\Lambda}\circ {\Gamma}_{\mu})$ and $\bar{f}_{B}(\bar{\Lambda}\circ {\Gamma}_{\mu})$, which are the average survive probabilities
of the modified quantum map $\bar{\Lambda}\circ {\Gamma}_{\mu}$,
\begin{eqnarray}
\chi_{\mu;\mu}(\bar{\Lambda})= \frac{1}{D^2}\{1+ (D+1)^2\bar{f}_{AB}(\bar{\Lambda}\circ {\Gamma}_{\mu}) \nonumber\\
- (D+1)[\bar{f}_{A}(\bar{\Lambda}\circ {\Gamma}_{\mu})+\bar{f}_{B}(\bar{\Lambda}\circ {\Gamma}_{\mu})]\}.
\end{eqnarray}
\begin{figure} \centering
\includegraphics[scale=0.3]{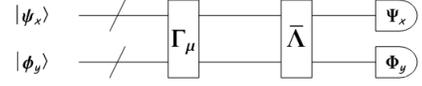}
\caption{\label{fig:epsart} Circuit representation of measuring the diagonal matrix element of the quantum  process matrix.}
\end{figure}

In FIG. 3, we give a protocol to measure the diagonal elements of the process matrix: Suppose that $\Psi_x$ and ${\Phi_y}$ are the elements of the state 2-design in (4). Preparing $\vert\psi_x\rangle \otimes\vert \phi_y\rangle$ as the input, perform an operation $\Gamma_{\mu}$ before the quantum
$\bar{\Lambda}$, then measure the expectations ${f}_{AB}(\bar{\Lambda}\circ \Gamma_{\mu})=\mathrm{Tr}[\Psi_x\otimes \Phi_y \bar{\Lambda}\circ \Gamma_{\mu}(\Psi_x\otimes \Phi_y)] $, ${f}_{A}(\bar{\Lambda}\circ \Gamma_{\mu})=\mathrm{Tr}[\Psi_x\otimes \mathrm{I}_D \bar{\Lambda}\circ \Gamma_{\mu}(\Psi_x\otimes \Phi_y)] $ and ${f}_{B}(\bar{\Lambda}\circ \Gamma_{\mu})=\mathrm{Tr}[\mathrm{I}_D\otimes \Phi_y \bar{\Lambda}\circ \Gamma_{\mu}(\Psi_x\otimes \Phi_y)] $. By repeating this process with a number of $N^2$ different inputs, one may get the averaged survive probabilities which can be  used to estimate the diagonal  element of the process matrix according to (52).

As it has been shown in [18], the way to decide the off-diagonal matrix elements is different form the one to get the diagonal matrix elements.
Let's organize the proposal developed  there in the way like: For the two known operators, $\Gamma_{\mu}$ and $\Gamma_{\nu}$, ($\mu\neq\nu$), define the following four operators,
\begin{equation}
\Gamma_{\pm}=\Gamma_{\mu}\pm\Gamma_{\nu}, \tilde{\Gamma}_{\pm}=\Gamma_{\mu}\pm i\Gamma_{\nu}.
\end{equation}
From it, one may easily verify that
\begin{eqnarray}
\Gamma_{\mu}\otimes\Gamma^*_{\nu}=\frac{1}{2}[\Gamma_+\otimes\Gamma_+^* -\Gamma_-\otimes\Gamma_-^*\nonumber\\
-i(\tilde{\Gamma}_+\otimes\tilde{\Gamma}_+^*- \tilde{\Gamma}_-\otimes\tilde{\Gamma}_-^* ], \nonumber\\
\Gamma_{\nu}\otimes \Gamma_{\mu}^*=\frac{1}{2}[\Gamma_+\otimes\Gamma_+^* -\Gamma_-\otimes\Gamma_-^*\nonumber\\
+i(\tilde{\Gamma}_+\otimes\tilde{\Gamma}_+^*- \tilde{\Gamma}_-\otimes\tilde{\Gamma}_-^* ]. \nonumber
\end{eqnarray}
Jointing this result with  (49), we observe that the off-diagonal  elements of the process matrix  can be estimated by the way
\begin{eqnarray}
\chi_{\mu;\nu}(\bar{\Lambda})=\frac{1}{2}[\omega_+-\omega_--i(\tilde{\omega}_+-\tilde{\omega}_-)],\nonumber\\
 \chi_{\nu;\mu}(\bar{\Lambda})=\frac{1}{2}[\omega_+-\omega_-+i(\tilde{\omega}_+-\tilde{\omega}_-)],\nonumber
\end{eqnarray}
in which the four quantities, $\omega_{\pm}$ and $\tilde{\omega}_{\pm}$, are defined as
\begin{eqnarray}
\omega_{\pm}=\frac{1}{D^4}\mathrm{Tr}[(\sum_{n}\bar{E}_n\otimes \bar{E}_{n})(\Gamma_{\pm}\otimes \Gamma_{\pm}^*)]\\
\tilde{\omega}_{\pm}=\frac{1}{D^4}\mathrm{Tr}[(\sum_{n}\bar{E}_n\otimes \bar{E}_{n})(\tilde{\Gamma}_{\pm}\otimes \tilde{\Gamma}_{\pm}^*)]
\end{eqnarray}

Let's consider $\omega_+$ at first. Following  the argument from (50) to (51), we can also interpret $\omega_+$ as the entanglement fidelity  of the
non-physical map $\bar{\Lambda}\circ \Gamma_+$
\[\omega_+=f^{\mathrm{ent}}(\bar{\Lambda}\circ \Gamma_+)\]
with the  process super operator $\bar{\Lambda}\circ \Gamma_+ $ defined  as $\bar{\Lambda} \circ {\Gamma}_{+}(\rho)=\sum_n \bar{E}_n(\Gamma_{+}\rho\Gamma_{+}^{\dagger})\bar{E}_n^{\dagger}$. With a simple calculation, which is  similar with the one for deriving (40), we shall get
\begin{eqnarray}
\omega_+=&& \frac{1}{D^2}\{2+ (D+1)^2\bar{f}_{AB}(\bar{\Lambda}\circ {\Gamma}_{+}) \nonumber\\
&-& (D+1)[\bar{f}_{A}(\bar{\Lambda}\circ {\Gamma}_{+})+\bar{f}_{B}(\bar{\Lambda}\circ {\Gamma}_{+})]\}.
\end{eqnarray}
In the derivation of it, we have used the relation, $\mathrm{Tr }[\vert \mathrm{I}^{\otimes 2}_D)( \mathrm{I}^{\otimes 2}_D\vert(\sum_{n}\bar{E}_n\otimes \bar{E}_{n})(\Gamma_{\pm}\otimes \Gamma_{\pm}^*)]=2D^2.$
Now, we shall encounter the problem of measuring the average survive probabilities for a non-physical map $\bar{\Lambda}\circ \Gamma_+$. A solution
for it, as it has been suggested in [18], is to prepare a set of states $\{\vert \psi_{xy}\rangle \}_{x,y=1}^{N}$,
\begin{equation}
\vert \psi_{xy}\rangle =\Gamma_+(\vert \psi_x\rangle\otimes \vert \phi_y\rangle),
\end{equation}
as the inputs for the quantum channel $\bar{\Lambda}$, then decide the averaged survived probabilities, which are needed in (56), through the way in below:
\begin{eqnarray}
\bar{f}_{AB}(\bar{\Lambda}\circ {\Gamma}_{+})&=&\frac{1}{N^2}\sum_{x,y=1}^{N}\mathrm{Tr}[(\Psi_x\otimes\Phi_y)\bar{\Lambda}(\Psi_{xy})], \nonumber\\
\bar{f}_{A}(\bar{\Lambda}\circ {\Gamma}_{+})&=&\frac{1}{N^2}\sum_{x,y=1}^{N}\mathrm{Tr}[(\Psi_x\otimes \mathrm{I}_D)\bar{\Lambda}(\Psi_{xy})],\nonumber\\
\bar{f}_{B}(\bar{\Lambda}\circ {\Gamma}_{+})&=&\frac{1}{N^2}\sum_{x,y=1}^{N}\mathrm{Tr}[(\mathrm{I}_D\otimes\Phi_y)\bar{\Lambda}(\Psi_{xy})].\nonumber
\end{eqnarray}
The product state 2-design protocol designed  to get the above average quantities is depicted in FIG. 4. The above method, which is developed for measuring $\omega_+$, can be easily generalized for the cases where the rest of the quantities defined in (54-55) should be measured.
\begin{figure} \centering
\includegraphics[scale=0.3]{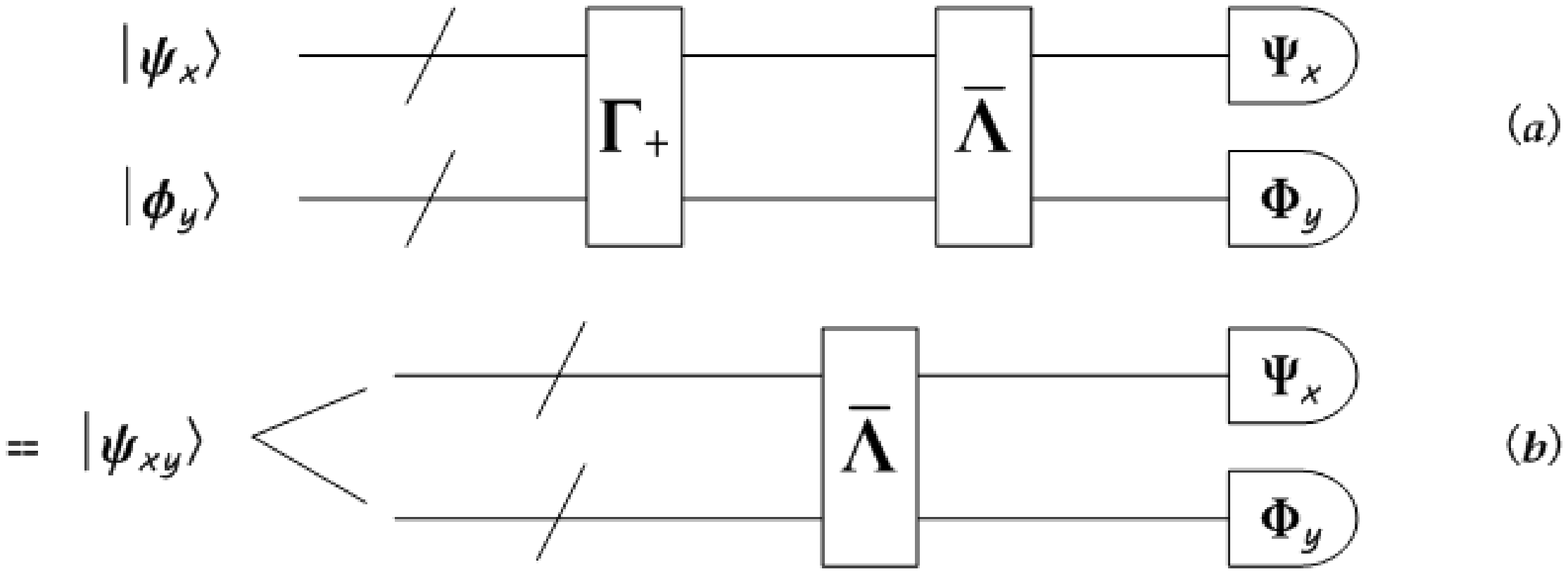}
\caption{\label{fig:epsart} (a) Measuring the survive probabilities of a non-physical map $\bar{\Lambda}\circ \Gamma_+$ can be realized by the apparatus in (b) with the inputs $\vert \psi_{xy}\rangle=\Gamma_+\vert \psi_x\rangle\otimes\vert\phi_y\rangle$.}
 \end{figure}

  \section{approximated measurement of the average survive probabilities}
In  section III, we have introduced three types  average survive probabilities for a bipartite system. Several ways of measuring these quantities are depicted in FIG. 2. It should be noted that these measurements are non-scalable: Taking the product state 2-design protocol, which is given in FIG. 2d, for an example,  there are  about $N^2 (N\ge D^2)$  runs of experiment  to be performed.   A solution for this problem is to find an efficient approximated way where the error introduced by the approximation should  depend on the number of operations   irrespective of the dimension of the joint system.

In this section, we shall develop such an efficient way to measure of the average survive probabilities. In what follows we restrict ourself to the case
where a set of product states, $\{\Psi_{i}\otimes \Psi_j\}_{i,j=1}^{M}$ should  be prepared as the inputs of the quantum channel $\Lambda$ while     the average survive probabilities defined in (19-21) are  approximated with the way like
\begin{eqnarray}
\bar{f}^{\mathrm{appr}}_{AB}=\frac{1}{M^2}\sum_{i,j,=1}^M\mathrm{Tr}[(\Psi_i\otimes \Psi_j)\Lambda(\Psi_i\otimes \Psi_j)],\nonumber\\
\bar{f}^{\mathrm{appr}}_{A}=\frac{1}{M^2}\sum_{i,j,=1}^M\mathrm{Tr}[(\Psi_i\otimes \mathrm{I}_{D})\Lambda(\Psi_i\otimes \Psi_j)],\nonumber\\
\bar{f}^{\mathrm{appr}}_{B}=\frac{1}{M^2}\sum_{i,j,=1}^M\mathrm{Tr}[(\mathrm{I}_D\otimes \Psi_j)\Lambda(\Psi_i\otimes \Psi_j)].\nonumber
\end{eqnarray}
These average  quantities can be also measured with the twirling procedures by letting $\Psi_j=U_j\Psi_0U_j^{\dagger}$ with $\Psi_0$ and $U_j$ to be a fixed state and  the arbitrary unitary transformations in $\mathrm{H}_{D}$, respectively.
 By following the steps for getting the super operators  in (33-35),  the above equations can be also expressed
\begin{equation}
  \bar{f}^{\mathrm{appr}}_{\alpha}=\mathrm{Tr}[\hat{F}^{\mathrm{appr}}_{\alpha}\lambda], ~~\alpha=A, B, AB.
  \end{equation}
in which the super operators $\hat{F}^{\mathrm{appr}}_{\alpha}$ are defined as
\begin{eqnarray}
\hat{F}^{\mathrm{appr}}_{A}=\beta(\frac{1}{M^2}\sum_{i,j=1}^{M}\vert\Psi_i\rangle\rangle\langle\langle \Psi_i\vert\otimes \vert\Psi_j\rangle\rangle\langle\langle \mathrm{I}_D\vert)\beta,\\
\hat{F}^{\mathrm{appr}}_{B}= \beta(\frac{1}{M^2}\sum_{i,j=1}^{M}\vert\Psi_i\rangle\rangle\langle\langle \mathrm{I}_D\vert\otimes \vert\Psi_j\rangle\rangle\langle\langle \Psi_j\vert)\beta,\\
\hat{F}^{\mathrm{appr}}_{AB}=\beta(\frac{1}{M^2}\sum_{i,j=1}^{M}\vert\Psi_i\rangle\rangle\langle\langle \Psi_i\vert\otimes \vert\Psi_j\rangle\rangle\langle\langle \Psi_j\vert)\beta.
\end{eqnarray}

 In above sections, we have shown that the average fidelity can expressed as the expectation of the super operator $\lambda$, which is defined
  in (31), with the separable Werner state $\tilde{\rho}^{\mathrm{sep}}_{\mathrm{W}}$, $f^{\mathrm{avg}}(\Lambda)=\mathrm{Tr}[\tilde{\rho}^{\mathrm{sep}}_{\mathrm{W}}\lambda]$.
  With the expanding formula of the identity operation in (37), one may rewrite $\tilde{\rho}^{\mathrm{sep}}_{\mathrm{W}}$ as
  \begin{equation}
  \tilde{\rho}^{\mathrm{sep}}_{\mathrm{W}}=\frac{2}{D^2(D^2+1)}\vert \mathrm{I}^{\otimes 2}_ D)(\mathrm{I}^{\otimes 2}_D\vert+\frac{1}{D^2+1}\Delta,
  \end{equation}
  in which the operator $\Delta$ is defined as
    \begin{eqnarray}
  \Delta&=& \frac{1}{D^2}(\mathrm{I}^{\otimes 4}_D- \vert \mathrm{I}^{\otimes 2}_ D)(\mathrm{I}^{\otimes 2}_D\vert)\nonumber\\
  &=& (D+1)^2\hat{F}_{AB}-(D+1)(\hat{F}_A+\hat{F}_B).
    \end{eqnarray}
  Now, the average fidelity is approximated in the way like
  \[f_{\mathrm{appr}}^{\mathrm{avg}}(\Lambda)=\mathrm{Tr}[\tilde{\rho}_{\mathrm{appr}}\lambda]\]
  with $\tilde{\rho}_{\mathrm{appr}}$ the approximated  separable Werner state,
  \begin{eqnarray}
  \tilde{\rho}_{\mathrm{appr}}=\frac{2}{D^2(D^2+1)}\vert \mathrm{I}^{\otimes 2}_ D)(\mathrm{I}^{\otimes 2}_D\vert+\frac{1}{D^2+1}\Delta_{\mathrm{appr}},\\
  \Delta_{\mathrm{appr}}=(D+1)^2\hat{F}^{\mathrm{appr}}_{AB}-(D+1)(\hat{F}^{\mathrm{appr}}_A+\hat{F}^{\mathrm{appr}}_B).
  \end{eqnarray}

Considering the fact that  the super operator $\lambda$, which  is  decided by the  quantum map $\Lambda$, is usually unknown,
 for simplicity, we suppose that the accuracy of the approximation  should  depend on the difference between the separable Werner state and its approximated form. As it has been done in [25], we  introduce  the Hilbert-Schmidt norm  of their difference,
 \begin{equation}
\vert \vert \tilde{\rho}^{\mathrm{sep}}_{\mathrm{W}}-\tilde{\rho}_{\mathrm{appr}}\vert\vert:=\mathrm{Tr}[(\tilde{\rho}^{\mathrm{sep}}_{\mathrm{W}}-
\tilde{\rho}_{\mathrm{appr}})
(\tilde{\rho}^{\mathrm{sep}}_{\mathrm{W}}-\tilde{\rho}_{\mathrm{appr}})^{\dagger}],
\end{equation}
 to characterize how well the separable Werner state
is approximated. With the definitions from (62) to (64), we simplify the above equation into the form
\begin{equation}
\vert \vert \tilde{\rho}^{\mathrm{sep}}_{\mathrm{W}}-\tilde{\rho}_{\mathrm{appr}}\vert\vert=\frac{1}{(D^2+1)^2}\vert\vert\Delta-\Delta_{\mathrm{appr}}\vert\vert.
\end{equation}

In order to perform the above calculation in an easy way, we shall introduce the  operators $R_{i}$,
\begin{equation}
R_{i}=(D+1)\Psi_i-\mathrm{I}_{D},
\end{equation}
and reexpress the operator   $\Delta_{\mathrm{appr}}$ in (65) with a more compact form
\begin{equation}
\Delta_{\mathrm{appr}}=\frac{1}{M^2}\sum_{i,j=1}^{M}\vert \Psi_i\otimes \Psi_j)(R_i\otimes R_j-\mathrm{I}^{\otimes 2}_D\vert.
\end{equation}
With above formula, one may easily verify that
\[\mathrm{Tr}[\Delta (\Delta)^{\dagger}]=\mathrm{Tr}[\Delta (\Delta)^{\dagger}_{\mathrm{appr}}]=\mathrm{Tr}[\Delta_{\mathrm{appr}} (\Delta)^{\dagger}]=\frac{2(D^2-1)}{D^2}.\]
Therefore, the Hilbert-Schmidt norm  is
\begin{equation}
\vert \vert \tilde{\rho}^{\mathrm{sep}}_{\mathrm{W}}-\tilde{\rho}_{\mathrm{appr}}\vert\vert=\frac{1}{(D^2+1)^2}(\vert\vert\Delta_{\mathrm{appr}}
\vert\vert-\frac{2(D^2-1)}{D^2})
\end{equation}
where the quantity $\vert\vert\Delta_{\mathrm{appr}}\vert\vert$ should be decided by our actual choice of the states $\Psi_i$.

Now, we assume that the states $\Psi_i$ are chosen from the symmetric information complete set (SIC) introduced in (6), $\Psi_i\neq \Psi_j$  for $i\neq j$. Based on this assumption,
we find that
   \begin{eqnarray}
    \vert\vert\Delta_{\mathrm{appr}}
\vert\vert&=& \frac{D^4(D+2)^2+(D^2-2)D^2}{M^2(1+D)^2}\nonumber\\
&& -\frac{4D}{M(1+D)}+\frac{D-1}{D+1}.
\end{eqnarray}
Putting it back into (70), we can get  the upper bound of the Hilbert-Schmidt norm,
\begin{equation}
\vert \vert \tilde{\rho}^{\mathrm{sep}}_{\mathrm{W}}-\tilde{\rho}_{\mathrm{appr}}\vert\vert\{\begin{array}{c}
                                                                                                =0,~~~~~~~~~~ M=D^2, \\
                                                                                              < \frac{1}{M^2}\frac{(1+D)^2}{D^2},~~1< M < D^2.
                                                                                              \end{array}
\end{equation}
For the case   $D\ge 4$, there is  $\frac{(1+D)^2}{D^2}<2$. Therefore, we conclude that  the separable Werner state is approximated in an efficient way: The upper bound  of the error introduced by the approximation process scales better than $\frac{2}{M^2}$ with the number of repetitions $M^2$ of the experiment.

\section{Discussion}
In previous works, the state 2-design has also  been defined in the way like:
\begin{equation}
\frac{1}{N}\sum_{x=1}^{N}\mathrm{Tr}[\Psi_x A \Psi_x B]=\frac{\mathrm{Tr}[A]\mathrm{Tr}[B]+\mathrm{Tr}[AB]}{D(D+1)}.
\end{equation}
One may check that the  definition  in (4) is consisted with it.
Applying the isomorphism  introduced in (9), we rewrite the left side of the above equation as
 \[\frac{1}{N}\sum_{x=1}^{N}\mathrm{Tr}[\Psi_x A \Psi_x B]
 =\mathrm{Tr}[(\frac{1}{N}\sum_{x=1}^{N} \Psi_x \otimes \Psi_x^*)(A\otimes B^{\mathrm{T}})].\]
As it has been shown in section II,  from  the definition of the state 2-design in (4), we can  get  an extremely simple relation between the state 2-design and the separable Werner state,
$\frac{1}{N}\sum_{x=1}^{N} \Psi_x \otimes \Psi_x^*=\rho^{\mathrm{sep}}_{\mathrm{\mathrm{W}}}.$
With the results, $\mathrm{Tr}[B]=\mathrm{Tr}[B^{\mathrm{T}}]$ and $(\mathrm{I}^{\otimes 2}_D\vert A\otimes B^{\mathrm{T}}\vert \mathrm{I}^{\otimes 2}_D)=\mathrm{Tr}[AB]$, we have
\[\mathrm{Tr}[\rho^{\mathrm{sep}}_{\mathrm{W}}(A\otimes B^*)]=\frac{\mathrm{Tr}[A]\mathrm{Tr}[B]+\mathrm{Tr}[AB]}{D(D+1)}.\]
From it, one may   conclude that the definition in (4) is consisted with the one in (73).

Finally, let's end our work with a short conclusion. For the bipartite system, we have introduced three directly measurable quantities, the average survive probability of the product states and the survive probability for each subsystem, and developed several protocols to measure them. These average quantities can be applied to estimate the average fidelity of the quantum channel and the selected element of the quantum process matrix.

\begin{appendix}
\emph{Appendix.} In Section II,  we have introduced a super operator $\hat{F}_b$,
$\hat{F}_b =\int d\mu_{\mathrm{H}}(U)U\otimes U^*\vert \Psi_0\rangle\rangle\langle\langle \Psi_0\vert (U\otimes U^*)^{\dagger}$,
and stated that it should equal with the separable Werner state defined in (17). Although this statement can be viewed as a known result
in other works, (for example, the one in [10]), for the completeness of present work, we shall still give a compact proof for it.
At first, applying the isomorphism in (22), we reexpress $\hat{F}_b$ in the vector form as $\vert \hat{F}_b )=\mathcal{T}\vert \Psi_0\rangle\rangle\otimes \vert \Psi^*_0\rangle\rangle$ with the super operator  $\mathcal{T}$ to be
\[  \mathcal{T}=\int d\mu_{\mathrm{H}}(U) U\otimes U^*\otimes U^*\otimes U.\]
In stead of directly carrying out the integration, we shall cite a result given by Scott in [26]:
\begin{eqnarray}
&&\int d\mu_{\mathrm{H}}(U)U\otimes U\otimes U^{\dagger}\otimes U^{\dagger}\nonumber\\
&&=\frac{1}{D^2-1}(P_{3412}+P_{4321})-\frac{1}{D(D^2-1)}(P_{4312}+P_{3421}),\nonumber
\end{eqnarray}
with $P_{abcd}$  defined as the permutation operator (For its definition in detail, please see the original work in [26].)
With the  above calculation in hands, we  find the expression of the  $\mathcal{T}$ should be
\begin{eqnarray}
\mathcal{T}=&&\frac{1}{D^2-1}[\vert \mathrm{I}_{D}^{\otimes 2})(\mathrm{I}_{D}^{\otimes 2}\vert- \vert \mathrm{I}_{D}^{\otimes 2})(\hat{S}_+\vert \nonumber \\
&&+D^2\vert \hat{S}_+)(\hat{S}_+\vert-\vert \hat{S}_+)(\mathrm{I}_{D}^{\otimes 2}\vert ],\nonumber
\end{eqnarray}
in which we use $\hat{S}_+$ to denote the projective operator, $\hat{S}_+=\vert S_+\rangle\langle S_+\vert$ with $\vert S_+\rangle$ the maximally entangled state  defined in (9). Performing   the operation $\mathcal{T}$ on an arbitrary product state $\vert \Psi_0\rangle \rangle\otimes \vert \Psi^*_0\rangle\rangle$,
we shall get
 \[ \mathcal{T}(\vert \Psi_0\rangle\rangle\otimes \vert \Psi^*_0\rangle\rangle)=\vert \rho^{\mathrm{sep}}_{\mathrm{W}}),\]
 the vector form of the desired  result $\hat{F}_b =\rho^{\mathrm{sep}}_{\mathrm{W}}$.

\end{appendix}


\begin{references}
\bibitem{1} M.A. Nielson, and I. L. Chuang, \emph{Quantum Computation and Quantum information}(Cambridge University Press, Cambridge,
UK.2000).
\bibitem{2} G. M. D'Ariano,  M. G.  A. Paris, and M. F. Sacchi,   Adv. Imaging. Electron. Phys. \textbf{128}, 205(2003).
\bibitem{3} G. M. D'Ariano  and J. LoPresti, in \emph{Quantum state
Estimation.} Edited by M. G.  A. Paris and J.\u{R}eh\'{a}\v{c}ek :
\emph{Lecture Notes in Physics}, Vol. \textbf{649 }(Springer, Berlin
2004).
\bibitem{4} I. L. Chuang  and  M. A. Nielson, J. Mod. Opt, \textbf{44}, 2455(1997).
\bibitem{5}  J. F. Poyatos, J. I. Cirac, and  P. Zoller,  Phys. Rev.
Lett. \textbf{78}, 390(1997).
\bibitem{6} D. W. Leung : Ph. D. thesis, Standford University, 2000;
\bibitem{7} G. M. D'Ariano , and J. LoPresti,    Phys. Rev. Lett. \textbf{86}, 4195(2001).
\bibitem {8}  J. B. Altepeter  \emph{et al.},   Phys. Rev. Lett. \textbf{90}, 193601(2003).
\bibitem{9} M. A. Nielsen, Phys. Lett. A \textbf{303}, 249(2002).
\bibitem{10} M. Horodecki, P. Horodecki, and R. Horodecki, Phys. Rev. A \textbf{60}, 1888(1999).
\bibitem{11} J. Emerson, R. Alicki, and K. \.{Z}yczkowski,   J. Opt. B \textbf{7}, S347(2005).
\bibitem{12}C. Dankert, R. Cleve, J. Emerson, and E. Livine, Phys. Rev. A \textbf{80}, 012304(2009).
\bibitem{13} J. Emerson, M. Silva, O. Moussa, C. Ryan, M.  Laforest, J. Baugh, D. G. Cory , and R.  Laflamme,  Science \textbf{317}, 1893(2007).
\bibitem{14} M. Silva, $\mathrm{Ph.D.}$ thesis, University of Waterloo, 2008.
\bibitem{15} C. C. L\'{o}pez, A. Bendersky, J. P. Paz, and D. G. Cory, Phys. Rev. A \textbf{81}, 062113(2010).
\bibitem{16} A. Bendersky, F. Pastawski, and J. P. Paz,  Phys. Rev. Lett.\textbf{100}, 190403(2008).
\bibitem{17} C. T.  Schmiegelow, M. A. Larotonda, and J. P. Paz,  Phys. Rev. Lett. \textbf{104}, 123601(2010).
\bibitem{18} C.T. Schmiegelow, A. Bendersky, M. A. Larotonda, and J. P. Paz, Phys.Rev.Lett. \textbf{107}, 100502(2011).
\bibitem{19} O. Moussa, M. P. da Silva, C. A. Ryan, and R. Laflamme, Phys. Rev. Lett. \textbf{109}, 070504(2012).
\bibitem{20}A. J. Scott,  J.Phys.A: Math. Gen. \textbf{39}, 13507(2006).
\bibitem{21}I. D. Ivanovi$\acute{c}$,  J.Phys.A: Math.Gen \textbf{14} 3241(1981).
\bibitem{22} W. K. Wootters and B. D. Fields, Ann. Phys. (N.Y.) \textbf{191}, 363(1989).
\bibitem{23}J. M. Renes, R. Blume-Kohout, A. J. Scott,  and C. M. Caves, J. Math.Phys. \textbf{ 45}, 2171(2004).
\bibitem{24} B. W. Schumacher, Phys. Rev. A \textbf{54}, 2614(1996).
     \bibitem{25} G. T\'{o}th and J. J. Garc\'{i}a-Ripoll, Phys. Rev. A \textbf{75}, 042311(2007).
     \bibitem{26}A. J. Scott,  J. Phys. A: Math.Theor. \textbf{41} 055308(2008).
\end{references}
\end{document}